\documentclass[twocolumn,showpacs,preprintnumbers,amsmath,amssymb]{revtex4}
\usepackage{graphicx}
\usepackage{dcolumn}
\usepackage{bm}
\begin{document}
\title{Phonon modulation of the spin-orbit interaction as a spin
relaxation mechanism in quantum dots}

\author{C. L. Romano}
\email[]{carlu@df.uba.ar}
\affiliation{Departamento de F\'{\i}sica, Universidad de Buenos Aires, C1428EHA Buenos Aires, Argentina}
\author{G. E. Marques}
\affiliation{Departamento de F\'{\i}sica, Universidade Federal de S\~ao Carlos, 13565-905 S\~ao Carlos SP, Brazil}
\author{L. Sanz}
\affiliation{Instituto de F\'{\i}sica, Universidade Federal de Uberl\^andia, 38400-902 Uberl\^andia MG, Brazil}
\author{A. M. Alcalde}
\affiliation{Instituto de F\'{\i}sica, Universidade Federal de Uberl\^andia, 38400-902 Uberl\^andia MG, Brazil}
\date{\today}

\begin{abstract}
We calculate the spin relaxation rates in a parabolic InSb quantum
dots due to the spin interaction with acoustical phonons. We
considered the deformation potential mechanism as the
dominant electron-phonon coupling in the Pavlov-Firsov spin-phonon
Hamiltonian.
By studying suitable choices of magnetic field and lateral dot size, we
determine regions where the spin relaxation rates can be practically suppressed.
We analyze the behavior of the spin relaxation rates as a function of
an external magnetic field and mean quantum dot radius. Effects of the spin admixture
due to Dresselhaus contribution to spin-orbit interaction are also
discussed.
\end{abstract}

\maketitle

\section{Introduction}

The ability to manipulate and control processes that involve transitions
between spin states is, at the moment, of extreme importance due to
the recent applications in polarized spin electronics and
quantum computation. Spin dephasing is the most critical aspect that should be
considered in the elaboration of proposals of quantum computation based
in single spin states as qubits in quantum dots (QDs)~\cite{Imamoglu99}.
Currently, QDs of diverse geometries are considered as promising candidates for
implementation of quantum computation devices because
the electronic, magnetic and optical properties can be controlled through
the modern grown and nanofabrication techniques. Due to the
long electron spin dephasing time
experimentally reported~\cite{Gotoh98,Gupta02,Paillard01}, the spin of an electron
localized in a QD has been suggested to realize a quantum bit.
While for bulk and for 2D systems
the spin relaxation processes have been studied in some
detail, the problem for QD's still require deeper and further discussions.
Several processes that can induce spin relaxation in semiconductors have been identified
and were studied.
At the moment remains in discussion which, between these processes, is dominant in
zero-dimensional systems.
Some experimental results have shown good agreement with the theoretical predictions for
2D systems~\cite{Lau01} but, in general, the identification of the processes
through direct comparison with the experimental results may become a formidable
task.
This problem is more critical for QDs, since few experimental results exist and the theoretical discussion
of the spin relaxation mechanisms is still an open subject.
Extensive theoretical works in QD systems have studied the
main phonon mediated spin-flip mechanisms, including admixture processes due to
spin-orbit coupling~\cite{Khaetskii01} and phonon coupling due to interface
motion (ripple mechanism)~\cite{Woods02}.
Spin relaxation rates depend strongly on the dot size, magnetic field strength, and
temperature, as
reported by several authors~\cite{Khaetskii01,Falko05}.
It was shown that the quantum
confinement produces, in general, a strong reduction of the QD relaxation rates.

In this work, we calculate the spin-flip transition rates,
considering the phonon modulation by the spin-orbit interaction.
For this purpose we will use the spin-phonon interaction
Hamiltonian proposed by Pavlov and
Firsov~\cite{Pavlov66,Pavlov67}. In this model, the Hamiltonian
describing the transitions with spin reversal, due to the
scattering of electrons by phonons, can be written in a general
form as
\begin{equation}
H_{ph}=V_\mathrm{ph} + \gamma[\sigma \times \nabla
V_\mathrm{ph}]\cdot (\mathbf{p}+e/c \mathbf{A}), \label{Rashba}
\end{equation}
where $V_\mathrm{ph}$ is the phonon operator, $\sigma$ is the Pauli spin operator, $\gamma$ is related with
the strength of the electron-phonon interaction,
$\mathbf{p}$ is the linear momentum operator and $\mathbf{A}$ is the vectorial potential related with
the external magnetic field $\mathbf{B}$.
This model has the advantage of being easily adapted to the
study of other interaction mechanisms with phonons.

\section{Theory}
Based on the effective mass theory applied to the problem of the
interaction of an electron with lattice vibrations, including the spin-orbit interaction
and in presence of an external magnetic field, Pavlov and Firsov~\cite{Pavlov66,Pavlov67} have obtained
the spin-phonon Hamiltonian that describes the transitions with spin reversal
of the conduction band electrons
due to scattering with longitudinal lattice vibrations as

\begin{widetext}
\begin{equation}
H_{ph} = d(\mathbf{q})\left( \frac{\hbar}{\rho_M V v q}\right)^{1/2}
\left\{ e^{i \mathbf{q \cdot r}} b_\mathbf{q}
\left[
\begin{array}{cc}
0 & \mathbf{\hat{n}}^-\times \mathbf{\hat{e}_q} \\
\mathbf{\hat{n}}^+\times \mathbf{\hat{e}_q} & 0
\end{array}
\right]
\left( \frac{\mathbf{p}}{\hbar} + \frac{e\mathbf{A}}{\hbar c} + \mathbf{q} \right)
+ \mathrm{h.c}\right\},
\label{spinphonon}
\end{equation}
\end{widetext}
where, $b_\mathbf{q} (b_\mathbf{q}^\dagger)$ are annihilation (creation) phonon operators,
the magnetic vector potential $\mathbf{A}$ is obtained in the
symmetric gauge considering an external magnetic field $\mathbf{B}$ oriented along the $z$ axis.
$\mathbf{\hat{n}}^\pm= \mathbf{\hat{x}} \pm i \mathbf{\hat{y}}$, where $\mathbf{\hat{x}}$, $\mathbf{\hat{y}}$
are unitary vectors along the $x$ and $y$ axis. $\mathbf{\hat{e}_q}$ is a unit vector in the direction
of the phonon polarization, $\mathbf{q}$ is
the phonon wave vector, $\mathbf{p}$ is the momentum operator, $v$ is the average sound
velocity, $\rho_M$ is the mass density, $V$ is the system volume and $d(\mathbf{q})$ is a
deformation potential electron-phonon coupling constant~\cite{Pavlov67}
\begin{equation}
d(q)=\frac{\hbar^2 q^2}{2m^\ast} \frac{E_0}{E_g}\left( 1-\frac{m^\ast}{m_0}\right)f\left(\frac{E_g}{\Delta}\right),
\end{equation}
where $E_g$=0.17eV is de band gap for InSb, $m^\ast$=0.013$m_0$,
$E_0$=7.0 eV is the deformation potential, $\Delta$=0.81 eV is the spin-orbit splitting of the
valence band, and $f(x)=(1+2x)/[(1+\frac{3}{2}x)(1+x)]$.

According to Eq.~(\ref{spinphonon}), we will restrict ourselves to the case of
\emph{bulk} phonon modes. The longitudinal acoustic phonons are described by
plane-waves of the form $\mathbf{u}_0 \mathbf{\hat{e}_q} \exp^{i
\mathbf{q \cdot r}}$, where $\mathbf{u}_0$ is the normalization
constant.
We consider the phonon energy dispersion in the Debye approximation: $E_q=\hbar v q$, for InSb we
use $v=3.4 \times 10^3$ m/s.
It is important to point out that the theoretical treatment
of the spin-phonon interaction developed by Pavlov and Firsov~\cite{Pavlov66,Pavlov67} can be easily
adapted to several electron-phonon interaction processes.
An appropriate choice of the electron-phonon coupling $d(\mathbf{q})$ allows us to adapt
the matrix element~(\ref{spinphonon}) for piezoacoustic or LO-optical spin scattering.

It has been assumed that the confinement along the $z$ axis is much stronger than the lateral confinement. Thus,
the lateral motion is decoupled from the one along $z$ and the envelope functions
separate $\psi(\mathbf{r})=f(x,y)\phi(z)$.
The $z$-dependent part of $\psi(\mathbf{r})$ is an eigenfunction of a symmetric quantum well of width $L$.
In lens-shaped quasi-two dimensional self
assembled QDs, the bound states of both electrons and
valence-band holes can be understood by assuming a lateral spatial confinement modeled by
a parabolic potential with rotational
symmetry in the $x-y$ plane \cite{Hawrylak99},
$V(\rho)=\frac{1}{2}m\omega_0^2\rho^2$, where $\hbar\omega_0$
is the characteristic confinement energy, and $\rho$ is the radial coordinate.
By using the one-band effective mass approximation and considering an
external magnetic field $B$ applied normal to the plane of the QD, the electron
lateral wave function can be written as
\begin{equation}
f_{n,M,\sigma}= C_{n,M}\frac{\rho^{|M|}}{a^{|M|+1}} e^{-\frac{\rho^2}{2a^2}} e^{iM\varphi}
L_n^{|M|}\left(\rho^2 / a^2 \right) \chi(\sigma),
\label{funciondeonda}
\end{equation}
where $C_{n,M}=\sqrt{n!/[{\pi(n + |M|)!}] }$, $L_n^{|M|}$ is the Laguerre polynomial, $n$ ($M$) is the principal (azimuthal)
quantum number, and $\chi(\sigma)$ is the spin wave function for the spin variable $\sigma$.
The corresponding eigenenergies are
\begin{equation}
E_{n,M,\sigma }=(2n+|M|+1)\hbar \Omega
+(M/2)\hbar \omega _{c}+(\sigma /2)g\mu _{B}B,
\end{equation}
where $\Omega =(\omega_{0}^{2}+\omega _{c}^{2}/4)^{1/2}$, $\mu _{B}$ is the Bohr magneton, $%
a=(\hbar /m\Omega )^{1/2}$ is the effective length, $\sigma = \pm 1 $ for spin up and down respectively, and $\omega _{c}=eB/m$.

In our model, we also consider the effects of the Dresselhaus contribution that provides
additional admixture between spin states. For 2D systems, the linear Dresselhaus Hamiltonian can be
written as
\begin{equation}
H_D=\frac{\beta}{\hbar}\left(\sigma_xp_x -\sigma_yp_y\right),
\label{dresselhaus}
\end{equation}
where $p_i = -i\hbar \nabla_i + (e/c)A_i$ with $i=x,y$ and $\beta$ is the Dresselhaus coupling parameter for this
confinement.
If the confinement potential in the $z$-direction
is considered highly symmetrical, then $\nabla V_z \sim 0$ and the Rashba contribution can
be safely ignored.

The spin relaxation rates ($W$) between the electronic states:
$(n,M,\uparrow (\downarrow)) \rightarrow (n^\prime, M^\prime, \downarrow (\uparrow))$, with
emission of one acoustic phonon, are calculated from the Fermi golden rule.

In the Hamiltonian (\ref{spinphonon}), we only consider the deformation potential
electron-phonon coupling, this is due to the large $g$-factor in narrow
gap InSb ($|g| \sim 51$), the dominant electron-phonon
coupling for spin relaxation is the deformation potential mechanism \cite{Alcalde04}.
The piezoelectric coupling governs the spin relaxation processes in wide or intermediate gap
semiconductors.
In the transition matrix elements calculation, we not only consider the linear
term $i \mathbf{q}\cdot \mathbf{r}$ in the
expansion of $\exp(i\mathbf{q} \cdot \mathbf{r})$~\cite{Khaetskii01}, but also the integral
representation of Bessel function is used in the evaluation of electron-phonon overlap integrals.
The linear approximation of $\exp(i\mathbf{q} \cdot \mathbf{r})$ may be valid
for spin inversion transitions in the spin polarized ground-states of GaAs based QDs where,
due to the small value of the electron $g$-factor, only long wavelength phonons are
involved.

\section{Results and discussion}
The calculations were performed for a parabolic InSb QD at $T\sim 0$~K. The material
parameters for the InSb system
are listed in Ref.~\cite{Destefani04a}.
We only have considered electron transitions between ground state electron Zeeman
levels $(0,0,\uparrow) \rightarrow (0,0,\downarrow)$ and
$(0,1,\downarrow) \rightarrow (0,1,\uparrow)$. The temperature dependence
for one-phonon emission rate is determined from $W=W_{0}(n_{B}+1)$, where $%
n_{B}$ is the Bose-Einstein distribution function and $W_{0}$ is the rate at $T=0$~K.
In the temperature regime $T \leq $10~K, we obtain $n_{B}+1\approx
1 $ and $W \approx W_{0}$. For temperatures larger than few Kelvin degrees ($T>10$~K),
two-phonon processes become the dominant spin relaxation
mechanism~\cite{Woods02}. These types of processes have not been considered in the present
calculation.

\begin{figure}[htp]
\includegraphics[width=18pc]{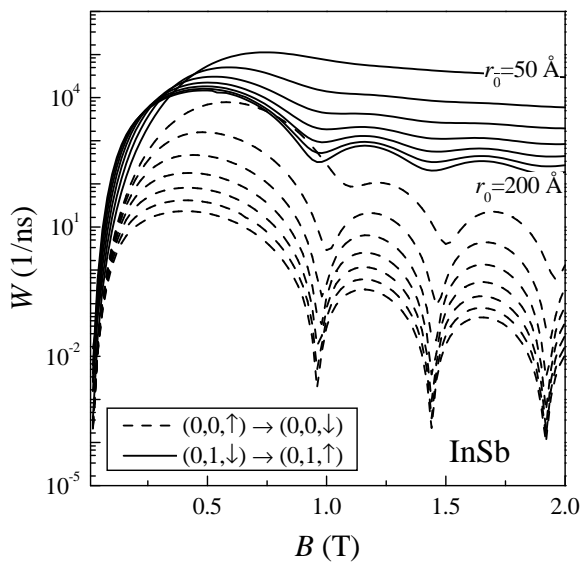}
\caption{(Color online) Spin relaxation rates, $W$, for a parabolic InSb QD considering the deformation potential
coupling mechanism as a function of the magnetic field $B$. We consider the
transitions: $(0,0,\uparrow) \rightarrow (0,0,\downarrow)$ and
$(0,1,\downarrow) \rightarrow (0,1,\uparrow)$ and
several lateral dot radius
$r_0$ = 50, 75, 100, 125, 150, 175, and 200 \AA (same $r_0$ ordering for both transitions) }
\label{rates1}
\end{figure}

\begin{figure}[t]
\begin{minipage}{18pc}
\includegraphics[width=20pc]{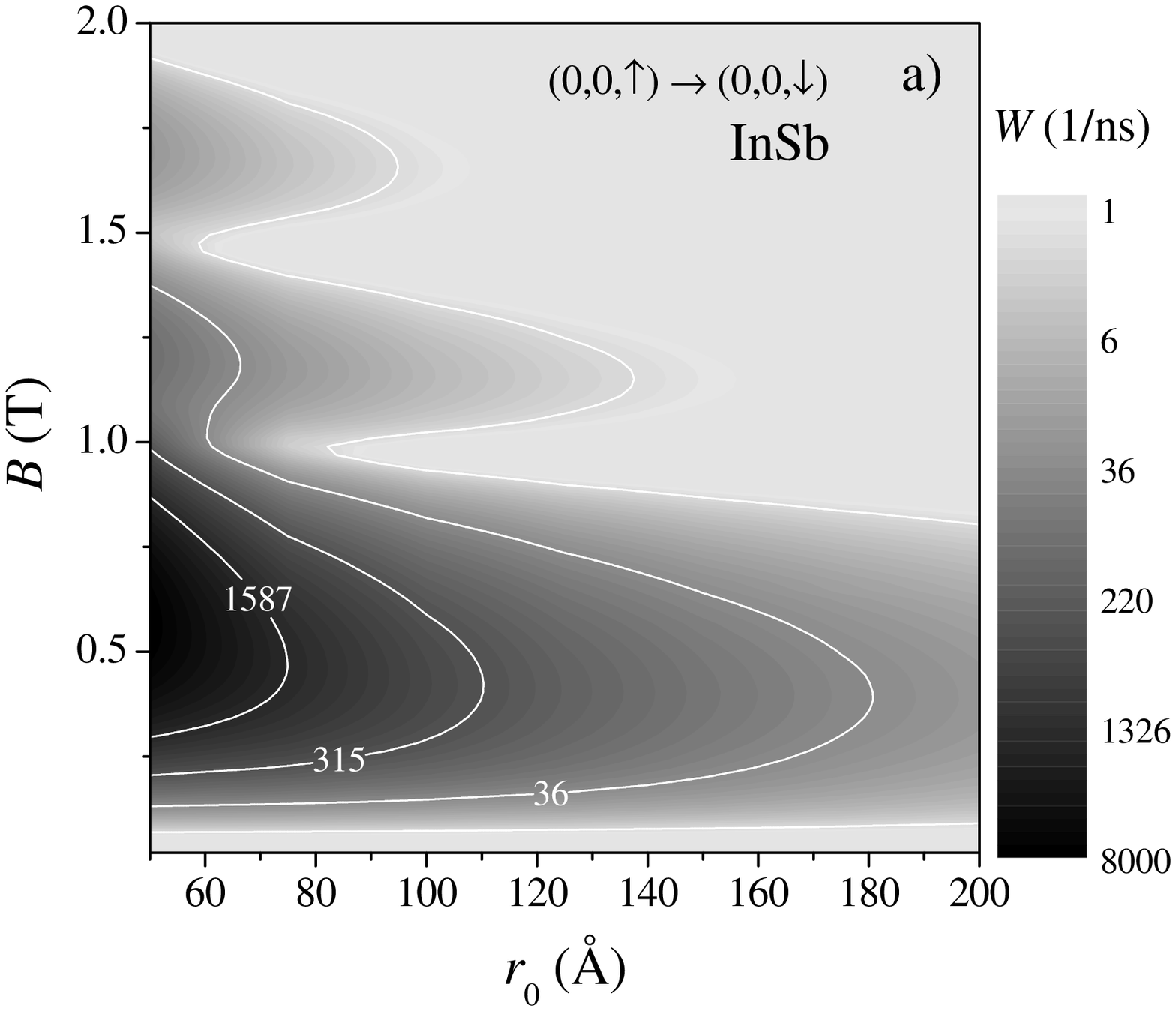}
\end{minipage}\hspace{2pc}%
\begin{minipage}{18pc}
\includegraphics[width=20pc]{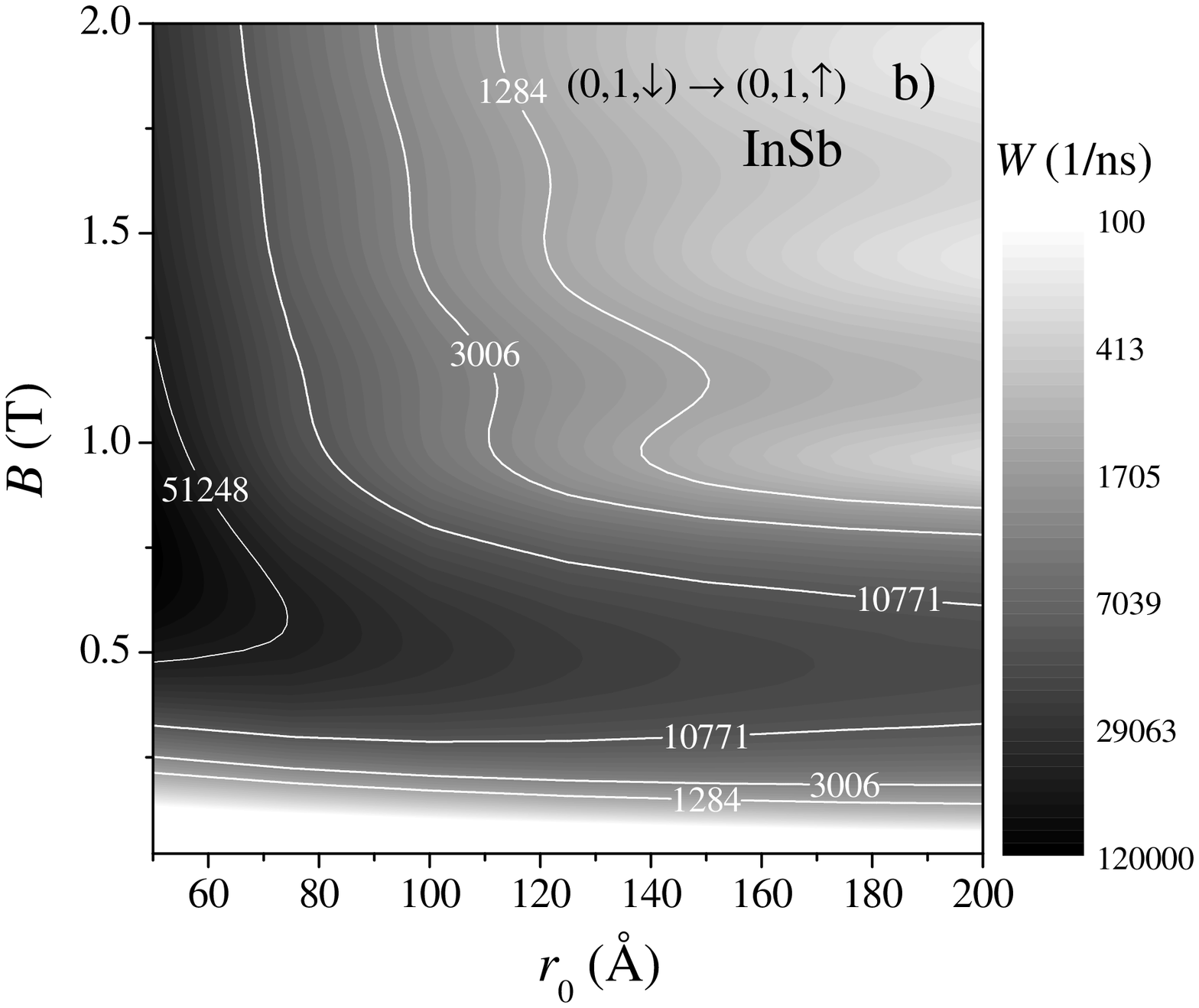}
\end{minipage}
\caption{(Color online) Contour plot of the spin relaxation rates as a function of magnetic field $B$ and
lateral size $r_0$ for
transitions: a) $(0,0,\uparrow) \rightarrow (0,0,\downarrow)$
and b)$(0,1,\downarrow) \rightarrow (0,1,\uparrow)$.}
\label{rates2}
\end{figure}

In the Fig.~\ref{rates1} we show the spin relaxation rates due to
deformation potential electron-phonon mechanism, as a function of the external
magnetic field $B$ and
considering some typical values for the effective lateral QD size,
$r_0=\sqrt{\hbar/m\omega_0}$.

Some interesting facts about these results should be
pointed out:

i) The rates show a strong dependence with the magnetic field.
This fact can be explained from the dependence of the rates
with the transition energy $\Delta E$.
In general, we obtain that $W \sim [g^\ast \mu_B B]^n=(\Delta E)^n$, $n$ being
an integer number that depends on the electron-phonon coupling process and $g^\ast$ the
effective $g$-factor.
As can be seen in Fig. \ref{rates1}, when the magnetic field increases, the rates also increase until
reaching a maximum near $B \sim 0.5$ T.
The position of this main maximum is defined from the transition energy conservation:
$E_{n l \sigma^\prime}-E_{n^\prime l^\prime \sigma^\prime}=\hbar v q$.

ii) The oscillatory behavior of the rates, observed for $B>0.7$T are mainly produced
by complex interplay between the spin admixture and electron-phonon overlap integrals.
The phonon modulated Rashba interaction, given in Eqs. (\ref{Rashba}) and (\ref{spinphonon})
produce level admixture according to the selection rules: $(n,M, \uparrow)\rightarrow (n,M+1,\downarrow )$,
$(n,M, \downarrow)\rightarrow (n,M-1,\uparrow )$
and $(n,M, \uparrow (\downarrow))\rightarrow (n,M,\downarrow (\uparrow) )$. The last condition, which
hybridize states with different spin orientations at the same level $M$, is the main responsible for the
rates oscillations at $B>0.7$T. For $\Delta M = 0$ transitions, the Dresselhaus SO interaction,
given in (\ref{dresselhaus}), is not
able to produce spin admixture.

As it is shown in Fig. \ref{rates1}, the $g^\ast$-factor effects are particulary important
for the ground-state Zeeman transition.
For small magnetic fields, $g^\ast \rightarrow g_\mathrm{bulk}$ and we may neglect the spin
admixture effects. Therefore, the spin relaxation shows no oscillations and
becomes almost independent of $r_0$.
This small QD size dependence is in agreement with the experimental
observations of Gupta and Kikkawa~\cite{Gupta99}.

iii) The rates dependence with the lateral
QD size $r_0$, are related to the interplay effects between the spatial and
magnetic confinements. This competing effects are contained in the electron-phonon overlap integral,
$I \propto \int f^\ast_{n^\prime,l^\prime,\sigma^\prime}(\rho) \exp(i\mathbf{q}\cdot\mathbf{r}) f_{n,l,\sigma}(\rho) d\mathbf{r}$.
For large fields, the magnetic confinement causes a gradual decrease in the overlap integral as the $r_0$ increases.
For small magnetic fields, the spatial confinement is dominant. Thus, when $r_0$ diminishes
the wave functions become more localized and the overlap integral should increase.
This effects explain the behavior of the spin transition $(0,1,\downarrow) \rightarrow (0,1,\uparrow)$ showed in
Fig.~\ref{rates1} (red lines). The Zeeman ground-state rates (black lines) are strongly
dependent on $\Delta E$ and, for small $B$, the rates are weakly dependent on $I$.

iv) The same rates calculated for GaAs (not showed here), are
in general, one order of magnitude smaller than InSb rates.
As we expected, the relaxation via piezoelectric coupling is more efficient than via the
deformation potential phonon processes.

In Fig. \ref{rates2} a) we have plotted the spin relaxation rates for the ground-state Zeeman
transition as a function of $r_0$ and $B$. We clearly identify a region of strong spin coherence,
defined by $B > 1$ T and $r_0 > 100$ \AA. In this regime, the relaxation times $\tau$ are in
the ns order and this is an important feature for spin qubit engineering.
In the $B < 0.1$T regime, the relaxation times are approximately of few $\mu$s.
This spin frozen
region is not robust against the temperature and will disappear whenever the thermal
energy is larger than the spin transition energy.
The plot in Fig. \ref{rates2} b) shows the spin rates for $(0,1,\downarrow) \rightarrow (0,1,\uparrow)$
transition. As in the previous case, the strong coherence regime is defined approximately
by $B > 1$ T and $r_0 > 100$ \AA. However, the relaxation times, in this region, are two orders of magnitude
faster than for $(0,0,\uparrow) \rightarrow (0,0,\downarrow)$ transition.

In conclusion, in this work we presented the calculation of spin relaxation
rates via electron-phonon interaction.
The model used in our work, allow us to estimate the magnitude of the scattering rates considering
the most important contributions to the spin relaxation through acoustic deformation potential interaction.
The rates exhibit a strong dependence with the magnetic field and with the lateral QD size.
We identify regimes of strong spin coherence, for magnetic field, $B> 1$T, and for lateral
sizes, $r_0 > 100$ \AA, we
obtain relaxation times approximately of 10$^{-2}$ ns to 1.0 ns.
The Zeeman ground state transition $(0,0,\uparrow) \rightarrow (0,0,\downarrow)$, exhibits
longer spin coherence times than the high energy transitions, which is important for
spin qubit applications.
The magnitudes of the calculated rates are compatible with those obtained experimentally.
For a more rigorous analysis of the spin relaxation process, many other effects should be
considered, like the anisotropy of the electron $g$-factor and
the effects of the spatial confinement on the acoustic phonons.
These two issues, are still under intense discussion and should be considered for a
more precise determination of the spin relaxation times.

\begin{acknowledgments}
This work has been supported by Funda\c{c}\~{a}o de Amparo \`{a} Pesquisa do
Estado de Minas Gerais (FAPEMIG) and by Conselho Nacional de Desenvolvimento
Cient\'{\i}fico e Tecnol\'{o}gico (CNPq).
\end{acknowledgments}


\end{document}